\documentstyle[12pt]{article}
 
\newcommand\beq{\begin{equation}}

\newcommand\eeq{\end{equation}}

\newcommand\parm{\par\medskip}

\newcommand\be{\begin{eqnarray}}
\newcommand\ee{\end{eqnarray}}

\def\np {Nucl. Phys.}
\def\prl {Phys. Rev. Lett.}
\def\pr {Phys. Rev.}
\def\pl {Phys. Lett.}

\def\bitem{\begin{itemize}}
\def\eitem{\end{itemize}}
\def\thefootnote{\fnsymbol{footnote}}

\begin{document}
\begin{flushright}
SUNY-NTG-92/27\\
\end{flushright}
\vskip .5cm
\vskip .25cm

\centerline{\large \bf Chiral Effective Action With Heavy Quark Symmetry
\footnote{Supported by the Department of Energy
under Grant No.\, DE-FG02-88ER40388. }}
\vskip 0.6in
\centerline{M.A. Nowak\footnote{Permanent address: Institute of Physics,
Jagellonian University, 30-059 Cracow, Poland.}
, M. Rho \footnote{Permanent address: Service de Physique
Th\'{e}orique, C.E. Saclay, 91191 Gif-sur-Yvette, France.} and I. Zahed}
\vskip 0.4in
\centerline{\it Department of Physics}
\centerline{\it State University of New York at Stony Brook}
\centerline{\it Stony Brook, New York 11794, USA}
\parm
\vskip 1.5 in

\centerline{\bf ABSTRACT}
\vskip .5cm

We derive an effective action combining chiral and heavy quark symmetry,
using approximate bosonization techniques of QCD. We explicitly show that
the heavy-quark limit is compatible with the large $N_c$ (number of color)
limit in the meson sector, and derive
specific couplings between the light and heavy mesons ($D$, $D^*$, ...)
and their chiral partners. The relevance of this effective action to
solitons with heavy quarks describing heavy baryons is discussed.

\par\vfill
\vfill
\noindent
\begin{flushleft}
SUNY-NTG-92/27\\ September 1992
\end{flushleft}
\eject

\newpage
\renewcommand{\thefootnote}{\arabic{footnote}}
\setcounter{footnote}{0}

\setcounter{equation}{0}
\setlength{\baselineskip}{27pt}
\noindent

The constraints of chiral symmetry on low energy processes have led to
a wealth of predictions ranging from strong to weak interactions. In
general, chiral symmetry constrains the dynamics of pions and kaons
by organizing the scattering amplitudes in powers of the light meson
momenta. Recently, it was suggested that chiral symmetry put also
constraints on the soft part of processes involving pions and heavy
mesons such as D and B \cite{wiseetal}.

It was suggested by Shuryak \cite{shuryak} and more recently
by Isgur and Wise \cite{isgurwise,references}
that if the mass of one quark is taken to infinity,
the dynamics of the heavy quark $Q$
is independent of its mass and spin. (We will refer to this limit as
Isgur-Wise limit or IW limit in short.) As a result, a new spin symmetry
develops in hadronic processes involving one heavy quark, leading to
a degeneracy of, say, D with D$^*$ and B with B$^*$. Several
relations \cite{isgurwise,references} have been recently derived showing
that the excitation
spectra and form factors are indeed independent of the mass and spin
of the heavy quark, a result analogous to the hydrogen atom.

The purpose of this letter is to provide a short derivation of the
effective action for QCD processes involving heavy and light
but nonstrange mesons, using approximate bosonization schemes,
a detailed discussion of which can be found in refs. \cite{bosoniz}.
In the presence of the light vector mesons our effective action differs
from the one presently in use in the litterature \cite{wiseetal}.
We show that the heavy-quark (IW) limit is compatible with the large
$N_c$ (number of color) limit in the meson sector,
and argue that heavy baryons $may$ be described as
solitons, as previously suggested \cite{heavyskyrme}
in the context of the Callan-Klebanov
model \cite{CK} and recently advocated by Manohar and
collaborators \cite{manoharetal}. While
our method could be readily extended to strange mesons as well, here
we shall restrict
our discussion to two light flavors $q=(u,d)$ and a heavy flavor ($Q$).
The heavy mesons ($\overline u Q$, $\overline d  Q$) transform as
singlets under chiral symmetry.

If the mass of the heavy quark is infinitely large, then the heavy
quark momentum is large and conserved $P_{\mu} = m_Q v_{\mu}$.
In this limit, there is a velocity superselection
rule \cite{georgi}. In the effective theory with heavy quarks,
this translates to
a different heavy quark (antiquark) field $Q_V^{\pm} (x)$ for each
velocity $v$. The latter carries momenta of the order of the
QCD scale $\Lambda$ and will be referred to as soft.
To display this we follow Georgi \cite{georgi} and define
\be
Q(x) = \frac{1+\rlap/v}2 e^{-im_Qv\cdot x} Q_v^+ (x)
       +\frac{1-\rlap/v}2 e^{im_Qv\cdot x} Q_v^- (x)\, .\label{e1}
\ee
As a result, the free QCD action reads
\be
S = \sum_v \int d^4x \,\, \left( \overline{q}(i\rlap/\partial - m_q ) q +
 \overline{Q}_v (i\rlap/v \,v\cdot\partial )Q_v\right)\, .\label{e2}
\ee
The action (\ref{e2}) is flavor $U(2)_L\times U(2)_R$ symmetric (for $m=0$)
and invariant
under independent spin rotations of the quark and the antiquark
(Isgur-Wise symmetry). The latter follows from the fact that the spin
effects are down by powers of $1/m_Q$. Finally, the decomposition
(\ref{e1}) is invariant under velocity shifts of the order of
$\Lambda$ -- a point recently stressed by Luke and Manohar \cite{lukeman}.
These conclusions are unaffected by the introduction of gluons to
leading order.

Approximate bosonization schemes for QCD have been discussed extensively
in the literature \cite{bosoniz}.
We will apply them here to the heavy-light system.
The idea consists of integrating out the short wavelength ($k>>\Lambda$)
content of the light quarks generating massive constituent quarks with
multiquark interactions (as in the instanton liquid model for instance)
admixed with bare but soft heavy quarks. In the long wavelength limit,
approximate bosonization schemes can be used to generate an effective
action as a gradient expansion in the slowly varying fields that
intermingles heavy-light dynamics.

Specifically, if we denote the heavy meson fields by
\be
\hat{H}_{\pm} = \frac{1+\rlap/v}{2}
(\gamma^{\mu}\hat{P}_{\mu,\pm}^{*}+i\gamma_5\hat{P_{\pm}} )
\gamma_5^{\pm}+\,\,{\rm h.c.}
 \label{e3}
\ee
where $\hat{P_+}^a\sim \overline{q}_R^a Q_v$,
$\hat{P}^{*a}_{\mu,+}\sim \overline{q}_R^a\gamma_{\mu} Q_v$ ($+\rightarrow -$
corresponds to $R\rightarrow L$)
are the pseudoscalar and vector {\em bare} heavy mesons with specific light
chirality, then standard arguments yield
\be
S =\sum_v \int \,\overline{\psi}\left(
{\bf 1}_2 i\rlap/\partial + {\bf 1}_3 i\rlap/v v\cdot\partial +
{\bf 1}_2 (\rlap/{\hat{L}}\gamma_5^+ + \rlap/{\hat{R}}\gamma_5^-)-
{\bf 1}_2 (M\gamma_5^+ + M^\dagger \gamma_5^-) +
\hat{H}_+ +\hat{H}_-\right) \psi
\label{action}
\ee
Here ${\hat{L}}_\mu\sim \overline{q}_L\gamma_{\mu} q_L$ and
${\hat{R}}_{\mu}\sim \overline{q}_R\gamma_{\mu} q_R$ are the bare light vector
fields, valued in $U(2)_L$ and $U(2)_R$ respectively,
${\bf 1}_2 = {\rm diag}(1,1,0),\,\,{\bf 1}_3 = {\rm diag}(0,0,1)$  are
the projectors onto the light and heavy sectors respectively and
we are using the short-hand notation $\gamma_5^\pm \equiv \frac{1}{2}(1\pm
\gamma_5)$.
The quark field $\psi$ in (\ref{action}) stands for $\psi =(q;Q_v)$.
The $\hat{P}$'s are off diagonal in flavor space, so that
$H= H_a T^a$ with $a=1,2$ and $T^1=(\lambda_4-i\lambda_5)/2,
\,\,T^2=(\lambda_6-i\lambda_7)/2$.

The action (\ref{action}) enjoys both heavy-quark spin symmetry
denoted $SU(2)_Q$
and rigid chiral  symmetry, $U(2)_L\times U(2)_R$.
We have set the light-quark masses
to zero and ignored the usual assumption-dependent part related to the
effective
potential in $M^\dagger M$. Here we will just assume that chiral
$U(2)_L\times U(2)_R$ is spontaneously broken to $SU(2)_V$ with the
appearance of three Goldstone bosons (the chiral anomaly taking care of the
$U(1)_A$) around a chiral symmetric condensate.
With this in mind, we will decompose the $2\times 2$
complex matrix $M$ as follows (a pertinent choice of gauge to avoid doubling
of the Goldstone bosons will be specified below)
\be
M=\xi_L^\dagger \Sigma\xi_R\label{decomp}
\ee
with the $\xi$'s as elements in the coset $SU(2)_L\times SU(2)_R /SU(2)_V$.
In the vacuum (saddle point) $\Sigma$ is diagonal and constant along the light
directions ($u,d$).
As a result, the ``bosonized" QCD action is in addition invariant under
local $SU(2)_V$ symmetry : $\xi_L\rightarrow h(x)\xi_L g_L^\dagger$,
$\xi_R\rightarrow h(x)\xi_R g_R^\dagger$ and
$\Sigma\rightarrow h(x)\Sigma h(x)^\dagger$.

The constituent (dressed) quark field $\chi$ relates to the bare quark
field $\psi$, through $\chi_{L,R} = (\xi_{L,R} q_{L,R}; Q_v )$. In terms of
the constituent field, the ``bosonized" action reads
\be
S = \sum_v\int d^4x&& \overline{\chi}  (
{\bf 1}_2 i\rlap/\partial + {\bf 1}_3 i \rlap/v v\cdot\partial +
{\bf 1}_2 (\xi_R i\rlap/\partial\xi_R^\dagger\gamma_5^+ +
           \xi_L i\rlap/\partial\xi_L^\dagger\gamma_5^- ) \nonumber\\
&& \left.  + {\bf 1}_2 (\xi_L \hat{\rlap/L}\xi_L^\dagger\gamma_5^+ +
            \xi_R  \hat{\rlap/R}\xi_R^\dagger\gamma_5^-)
 -  {\bf 1}_2 (\Sigma\gamma_5^+ + \Sigma^+\gamma_5^-)\right. \nonumber \\
&&  + \overline{H}+ H +\overline{G} +G ) \chi.
\label{action2}
\ee
It is invariant under local $SU(2)_V$ symmetry and global $SU(2)_Q$
symmetry. Now, let us define the dressed fields
\be
L_{\mu} &=&\xi_L\hat{L}_{\mu}\xi_L^\dagger + i\xi_L\partial_{\mu}
\xi_L^\dagger, \nonumber\\
R_{\mu} &=&\xi_R\hat{R}_{\mu}\xi_R^\dagger + i\xi_R\partial_{\mu}
\xi_R^\dagger,
\ee
with
\be
H&=& \frac{1+\rlap/v}2 (\gamma^{\mu}P^*_{\mu} +i\gamma_5P) =
     \frac{1+\rlap/v}2 (\gamma^{\mu}(P^*_{\mu,+}\xi_R^+ + P^*_{\mu,-}\xi_L^+)
                           +i\gamma_5(P_+\xi_R^+ +P_-\xi_L^+)),\nonumber\\
G&=& \frac{1+\rlap/v}2 (\gamma^{\mu}\gamma_5 Q^*_{\mu} +Q) =
     \frac{1+\rlap/v}2 (\gamma^{\mu}\gamma_5
(P^*_{\mu,+}\xi_R^+ - P^*_{\mu,-}\xi_L^+) +(P_+\xi_R^+ -P_-\xi_L^+))
\ee
where the new $H$ and $G$ fields refer to $(D,D^*)$ and their
{\it chiral partners} $(\tilde{D}, \tilde{D}^*)$
respectively. The heavy vector fields are transverse $v^{\mu}P^*_{\mu}
=v^{\mu}Q_{\mu}^*=0$. The ``bosonized" QCD action in the dressed fields becomes
\be
S=\sum_v\int d^4x\overline{\chi}\left(
{\bf 1}_2 (i\rlap/\nabla_L -\Sigma )\gamma_5^- +
{\bf 1}_2 (i\rlap/\nabla_R -\Sigma ) \gamma_5^++
{\bf 1}_3 i\rlap/v v\cdot\partial +
H+\overline{H}+ G+ \overline{G} \right)\chi \label{action3}
\ee
where the covariant $L,R$ derivatives are :
$\nabla_L = \partial -iL $ and $\nabla_R = \partial -iR $. $\overline{H}$ and
$\overline{G}$ are related, respectively, to $H$ and $G$ through
\be
\overline{H} =\gamma^0 H^+ \gamma^0\, ,\qquad\qquad
\overline{G} =\gamma^0 G^+ \gamma^0\, .
\ee

The light-light and heavy-light
quark dynamics follows from the dressed action (\ref{action3})
through a derivative
expansion. The momenta are bounded from above by a number times $\Lambda$
and from below by a soft quark mass $\epsilon\sim \Lambda^2/m_Q$ acquired
by the long wavelength components of the heavy-quark propagator,
if we recall that the heavy-quark condensate vanishes as $\Lambda^4/m_Q$.
This effect is dominant in the infrared regime for the soft part of
the heavy-quark field $Q_v$.
Our expansion of (\ref{action3}) will be understood in the sense of
$m_Q / \Lambda\rightarrow \infty$.

To second order, the heavy-light induced action reads
\be
S_H = &&N_c {\rm Tr}\left(
{\bf 1}_2 \Delta_l H {\bf 1}_3 \Delta_h \overline{H}\right) -\nonumber\\
&&N_c {\rm Tr}\left(
{\bf 1}_2 \Delta_l(\rlap/V \Delta_l +\rlap/A \Delta_l \gamma_5)
H {\bf 1}_3 \Delta_h \overline{H}\right) + \cdots
\label{actionh}
\ee
where the functional trace includes tracing over
space, flavor and spin indices.
We have denoted $\Delta_l = (i\rlap/\partial -\Sigma)^{-1}$
and $\Delta_h = (i\rlap/v v\cdot\partial)^{-1}$, and
\be
\gamma_5^+\rlap/R +\gamma_5^-\rlap/L = \rlap/V +\gamma_5\rlap/A\, .
\ee
The ellipsis in (\ref{actionh}) stands for higher insertions of
vectors and axials.
A similar action is expected for the heavy chiral partners $G$'s.
To the order considered, there are also cross terms generated by the axial
current.

After carrying out the trace over  space in (\ref{action}) and renormalizing
the heavy-quark fields, $H\rightarrow H/ \sqrt{Z_H}$ and
$G\rightarrow G/ \sqrt{Z_G}$, we obtain to leading order in the
gradient expansion for the $H$'s ($s_l^{P_l}={\frac 12}^-$)
\be
{\cal L}^H_v= &&-\frac i2 {\rm Tr}(\bar{H}v^{\mu}\partial_{\mu}H -
v^{\mu}\partial_{\mu}\bar{H}\,H)
\nonumber \\
&&+{\rm Tr}V_{\mu}\bar{H}Hv^{\mu} -g_H {\rm Tr} A_{\mu}\gamma^{\mu} \gamma_5
\bar{H} H +m_H {\rm Tr} \bar{H}H
\label{mano}
\ee
and for the $G$'s ($s_l^{P_l}={\frac 12}^+$)
\footnote{We label $H$ and $G$ by the total
angular momentum $s_l$ and parity $P_l$ of the {\em light} degrees of freedom.}
\be
{\cal L}^G_v= &&+\frac i2 {\rm Tr}(\bar{G}v^{\mu}\partial_{\mu}G -
v^{\mu}\partial_{\mu}\bar{G}\,G)
\nonumber \\
&&-{\rm Tr}V_{\mu}\bar{G}Gv^{\mu} -g_G {\rm Tr} A_{\mu}\gamma^{\mu} \gamma_5
\bar{G} G +m_G {\rm Tr} \bar{G}G
\label{manoh1}
\ee
The parameters in (13,14) are given by ($P_l=\pm$ is the parity of the light
part of $H,G$)
 \be
Z_{P_l}&=& N_c \int_{\epsilon}^{\Lambda}\frac{d^4Q}{(2\pi)^4}
            \frac{-1 /(v\cdot Q) +P_l\,2\Sigma /(Q^2-\Sigma^2)}{(Q^2-\Sigma^2)}
\nonumber\\
g_{P_l}&=& \frac {N_c}{Z_{P_l}}\,\,\,
 \int_{\epsilon}^{\Lambda}\frac{d^4Q}{(2\pi)^4}
             \frac{1}{v\cdot Q} \frac{-Q^2/3 +\Sigma^2}{(Q^2-\Sigma^2)^2}
\rightarrow 1 \nonumber\\
m_{P_l}&=&\frac {N_c}{Z_{P_l}}\,\,\,
\int_{\epsilon}^{\Lambda}\frac{d^4Q}{(2\pi)^4}
\frac{1 +P_l \,\Sigma /(v\cdot Q)}{Q^2-\Sigma^2}\rightarrow -P_l\,\Sigma
\label{para}
\ee
with $\epsilon\sim \Lambda^2/m_Q$ and the limits in (\ref{para}) follow from
$m_Q /\Lambda\rightarrow\infty$. Note that $m_H$ and $m_G$ are of order
$m_Q^0$. The interaction between the $H$'s and the
$G$'s is given by
\be
{\cal L}^{HG}_v = -\sqrt{\frac{Z_H}{Z_G}}{\rm Tr}
(\gamma_5\overline{G}H v^{\mu}A_{\mu})
-\sqrt{\frac{Z_G}{Z_H}}{\rm Tr}
(\gamma_5\overline{H}G \gamma^{\mu}A_{\mu})
\ee
Notice that $Z_G/Z_H = g_H/ g_G$ reduces to 1 in the heavy quark limit.
We observe that the mass splittings between the $H$'s ($D,D^*,...$)
and their chiral partners the $G$'s ($\tilde{D},\tilde{D}^*,...$)
imply the following mass relations to order $m_Q^0 N_c^0$
\be
m(\tilde{D}^*) -m(D^*) = m(\tilde{D}) -m(D) = \Sigma
\label{mass}
\ee
where again $\Sigma$ is the constituent  mass of the light quarks
in the chiral limit.
This is expected since the $D$, $D^*$ are S-wave mesons, while the
$\tilde{D}$, $\tilde{D}^*$ are P-wave mesons (not yet observed)\footnote{The
observed
$D_1(2400)$ and $D_2^*(2460)$ are components of the $s_l^{P_l}=\frac{3}{2}^+$
multiplet.}.  In the nonrelativistic
quark model the difference is centrifugal and of order  $m_Q^0$. This point
has been  appreciated already by Shuryak  in the context of
bag models and QCD sum rules \cite{shuryak}.

Since the
decomposition (\ref{decomp}) doubles the scalar degrees of freedom, a proper
gauge fixing in $\xi$'s is required. We choose the ``unitary gauge"
$\xi_L^{\dagger} =\xi_R\equiv \xi =e^{i\pi /2f_{\pi}}$.
In a minimal model with $only$ pions (model I), we have for the vector and
axial currents
\be
V_\mu = &&+ \frac i2 \left( \xi\partial_\mu\xi^\dagger+
\xi^{\dagger}\partial_\mu\xi \right),  \nonumber\\
A_\mu = && + \frac i2 \left( \xi\partial_\mu\xi^\dagger -
\xi^{\dagger}\partial_\mu\xi\right).
\ee
If we were to introduce vector mesons ($\omega, \rho, a_1(1270)$)
(model II) then the vector and axial
currents are $entirely$ vectorial
\be
V_\mu = && \omega_{\mu} +\rho_{\mu}\nonumber\\
A_\mu = &&  {\cal A}_{\mu} - \frac{\alpha}{f_{\pi}} \partial_{\mu}\pi
\label{model2}
\ee
where $\alpha = (m_{\rho}/m_A)^2$ follows after eliminating the
$\pi a_1$ mixing at tree level in the light-light sector \cite{bosoniz}.
The relations (\ref{model2}) identify the vector fields with
the constituent vector currents, $e.g.$
$\omega_{\mu}\sim \overline{\chi}\gamma_{\mu}\chi$,
as can be checked explicitly in (\ref{action3})
\footnote{Note that in this case the soliton approach to heavy hyperons
follows from binding $Q,Q^*$ instead of $P,P^*$ to Skyrmions.}. They also
enforce vector dominance in the light-light sector. Clearly other
constructions are also theoretically possible in which the amount of
pion dressing is intermediate between model I and model II, using the
relations (7). Of course, the issue of how the effective fields are
physically defined  (dressed) can only be resolved by comparing
the various model predictions with experiment.

The renormalized effective action involving both $H$ and $G$ and their
interactions is invariant under local $SU(2)_V$ (or more precisely
$U(2)_V$ including the singlet) symmetry (h), in which
$V$ transforms as a gauge field, $A$ tranforms covariantly and
$H,G \rightarrow H h^{\dagger},G h^{\dagger}$ and
$\overline{H},\overline{G}\rightarrow h\overline{H},h\overline{G}$.
It is also invariant under heavy-quark symmetry $SU(2)_Q$ (S),
$H,G\rightarrow SH,SG$ and $\overline{H},\overline{G} \rightarrow
\overline{H}S^{\dagger},\overline{G}S^{\dagger}$.

Our renormalized effective action (\ref{mano}) with only pions (model I)
is entirely consistent with the one suggested by Wise and others
(\cite{wiseetal})
(aside from the mass term). Our derivation suggests $g_H=1$ with
a specific $sign$ assignment for the axial term. The  magnitude
of $g_H$ is consistent with the constituent quark model.
The rest of our effective action is also consistent with the
effective action writen down recently by \cite{KORNER}.
\footnote{In the convention of \cite{FALK,KORNER} the P-wave heavy quark
field $H_{\alpha}$ relates to our $G$ through
$H_{\alpha}=G(v_{\alpha}-\gamma_{\alpha})/\sqrt{3}$.}

The introduction of vector mesons along with pions (model II) which is a more
realistic description of the light-light sector yields a totally $new$
effective action for the heavy-light sector. The identification of
the currents is entirely vectorial in this case. The pion coupling
to the heavy particles occurs $solely$ through the longitudinal
component of the axial current. In this case the $\pi$-HH coupling  is
quenched in comparison with the $a_1$-HH coupling ($1$ in this case),
$g_{\pi HH} = (m_{\rho}/m_A)^2 = .37$, improving
the decay width $D^*\rightarrow D\pi$ of model I by about $1/2$,
in the limit where the
heavy quark mass is infinite.\footnote{The recent CLEO collaboration
data seem to indicate a somewhat smaller empirical value $g_H \approx
.58$ with a large error bar. See \cite{CLEO} for an analysis.}

Our arguments suggest that in the heavy-quark limit
and in the chiral limit, a similar effective action should involve the
chiral partners of the heavy pseudoscalars and vectors, here denoted
by $G$. Their role in the soliton scenario might be important for the
description of opposite-parity heavy-baryon states. They also
allow for a qualitative estimation for the coefficients involved,
and provide a rationale for the systematic expansion in both
$k_{\pi}/(\sqrt{N_c}\Lambda)$ (derivative or $1/N_c$ expansion) and
$\Lambda/m_Q$
(heavy-quark expansion).

The scaling with $N_c$ of the overall heavy-light effective action before
renormalization, suggests that the heavy-light system
should be entrusted with the
same weight as the light-light counterpart (which is well known and hence
omitted in our discussion) in the large $N_c$ limit, implying that
a soliton description for heavy baryons is perhaps justified
\cite{heavyskyrme}. The large $N_c$ limit appears to be compatible with
the heavy-quark limit in the meson sector.
This point is $a$ $priori$ not obvious since
terms of the form $m_Q/N_c\Lambda$ and others cannot be ruled out on
general grounds. Thus our approximate bosonization version of QCD
yields a long wavelength description that appears to be
consistent with the one advocated recently by \cite{wiseetal}.
This is to be contrasted with the ``colored" mesonic
description recently discussed by Ellis $et$ $al$ \cite{ellisetal}
where it was argued
that in $QCD_2$ the presence of a large current quark mass forces
the bosonization to be $U(N_fN_c)\times U(N_fN_c)$ ({\it i.e.}, colored
bosons) as opposed to the factorized bosonization scheme
$(U(N_f)\times U(N_f))$$\otimes$$(U(N_c)\times U(N_c))$ (colorless bosons).
Here $N_f$ is the number of flavours.
It would be interesting to see how the approximate bosonization scheme
discussed here works in QCD$_2$ in comparison to the exact solutions
to the 't Hooft equations for heavy-light systems \cite{grinstein}.
A similar comparison
is also warranted for the solutions discussed by Ellis $et$ $al$.

We  note that that in the meson sector the $HH\pi$ interaction is
of order $1/\sqrt{N_c}$ as expected. In the soliton sector, however, the
effect of this interaction is in $principle$ of order $N_c$,
since we expect $H\sim \sqrt{N_c}$ and $\pi\sim\sqrt{N_c}$
({\it i.e.}, semiclassical field).
However, in the heavy-quark limit, the heavy meson
field is $in$ $general$ localized over a range of the order of
$1/(N_c m_Q)$ affecting
the energy to order $N_c^0$. Hence, one would expect heavy baryons
composed of pions and $P, P^*$ (or $Q, Q^*$) to emerge to order $N_c^0$ or
 lower.\footnote{The sharp localisation of
 the heavy quark field $H$ in the
heavy mass, large $N_c$ limit may be at odd with the soft character
of $H$ and thus invalidate {\em even further} the use of the derivative
expansion in the soliton sector of heavy baryons than in that of light-quark
baryons. (The derivative expansion for skyrmions in the light-quark sector
is bad for the same reason that the standard chiral perturbation
theory does not work with baryons and pions.)
Also there might be problems with the
commutativity of the heavy mass, large $N_c$ limit in the same sector.
These points are  presently under investigations.}
The emergence of the heavy baryon spectrum depends crucially on
whether the $P, P^*$ bind to the soliton to order $N_c^0$.

In the Callan-Klebanov scheme, the presence of the
Wess-Zumino term causes P-wave kaons to bind  to the soliton
(here $U_0 = \xi^2_0$)  to order $N_c^0$.
The bound state carries  good grand spin $K=I+J$ (here $1/2^-$) and
heavy-flavor
quantum number. However
states with good isospin ($I$) and angular momentum ($J$) emerge after
``cranking" (or rotating) the kaon-soliton bound state as a $whole$.
This means that
heavy meson soliton states with good quantum numbers are expected only
to order $1/N_c$. In the scheme recently advocated by Manohar and
collaborators\cite{manoharetal}, the heavy mesons are treated as free
wavepackets
of the size of the Compton wavelength of the heavy quark. In this case
{\em only} the soliton is cranked leaving the heavy meson field unrotated
and good quantum states composed of the soliton and the heavy meson appear
already to order $N_c^0$. At this order, all the baryons of flavor $Q$
({\it i.e.}, $\Lambda$, $\Sigma$, $\Sigma^*$ etc.) are degenerate.
The splitting $\Sigma-\Lambda$ occurs to order $1/N_c$,
while $\Sigma^*-\Sigma$ occurs to order $1/m_QN_c$. In this scheme,
the Wess-Zumino term plays no role.
In both approaches, good quantum numbers {\em and} fine-structure (or
$\Sigma-\Lambda$) splitting are obtained to order $1/N_c$. In the latter case,
the Isgur-Wise symmetry is automatic to that order but in the former case,
terms involving vector mesons have to be added to restore the IW symmetry.

An interesting question to ask is: Is it possible to interpolate
the Callan-Klebanov model to the heavy-quark limit?
We have no clear answer but we can think of two possibilities.
\bitem
\item As noted above, in the heavy-quark limit, the heavy mesons
decouple from the Wess-Zumino term.
Therefore, the primary source of binding in the
conventional form of the Callan-Klebanov model would be lost.
If there is no other source of binding, then the bound state
would disappear into the continuum for a finite but heavy-quark mass
(following the disappearance of the Wess-Zumino term),
causing the decoupling of the heavy kaon from the soliton to order $N_c^0$.
This would signal the breakdown of the Callan-Klebanov picture at some
large quark mass and the only viable heavy skyrmion would be the one
advocated in \cite{manoharetal}.
\item Although the topological Wess-Zumino term does not survive
heavy quark mass, the binding could still
persist all the way to the heavy-quark
limit. In fact our effective action with the inclusion of the omega
yields naturally a coupling of the form
${\rm Tr} (\overline H H )\,v\cdot\omega$
between the heavy meson and the omega (model II). If we recall that the
light-light
effective action induces a pion-omega coupling of the form
$N_c \omega^{\mu}B_{\mu}$, where
$B_{\mu}$ is the properly normalized baryon current, then by eliminating the
$\omega^0$ part through its constraint equation, we generate a term in the
Hamiltonian density of the form (heavy omega)
\be
{\cal H} \sim \frac {g^2}{2m_{\omega}^2} (N_c B^0 +v^0{\rm Tr} (\overline H
H))^2
\ee
Here $g\sim 1/\sqrt{N_c}$ is the vector gauge coupling. With our conventions,
the omega-three-pion coupling is
$g_{\omega} = N_c g\sim \sqrt{N_c}$. Clearly the choice
${\rm Tr}(\overline H H)v^0\sim - B^0$,
can lower the energy to order $N_c^0$.  This argument is similar to the one
advanced by \cite{min}. We see no $a$ $priori$ reason for suppressing the
heavy meson couplings to the light vector mesons.
In this case it would appear that the Callan-Klebanov description
would go {\it smoothly} over to the
the IW limit, provided the appropriate vector mesons are introduced
to restore the IW symmetry (see also footnote 6).
Approaching the IW limit in this way is
not perhaps very elegant but what is surprising is that it can be done at all.
\eitem
\noindent
Which of the two pictures is realized in Nature is an intriguing question
which we hope to answer.

\vskip 1cm
{\bf Acknowledgments}
\vskip .4cm

This work has been supported in part by a DOE grant DE-FG02-88ER40388
and by KBN grant PB 2675/2. We are grateful for comments from
J. Milana, D.-P. Min, N.N. Scoccola, E. Shuryak and J.J.M. Verbaarschot.

\pagebreak

\end{document}